\documentclass[a4paper,11pt]{article}
\pdfoutput=1
\usepackage{jheppub}
\usepackage{multirow}
\usepackage{amsmath,amssymb}
\usepackage{bbm,marvosym}
\usepackage{color}
\usepackage{graphicx}
\usepackage{enumerate}

\title{Anatomy of flavour-changing Z couplings in models with partial compositeness}
\author{David M.\ Straub}
\affiliation{
PRISMA Cluster of Excellence \& Mainz Institute for Theoretical Physics\\
Johannes Gutenberg University, 55099 Mainz, Germany}
\emailAdd{david.straub@uni-mainz.de}
\preprint{MITP/13-014}

\abstract{In models with partially composite quarks, like composite Higgs models or models with a warped extra dimension, the couplings of quarks to the $Z$ boson generically receive non-universal corrections that are not only constrained by electroweak precision tests but also lead to flavour-changing neutral currents at tree level.
The impact of these flavour-changing couplings on rare $K$ and $B$ decays is studied in two-site models for three scenarios: an anarchic strong sector with two different choices of fermion representations both leading to a custodial protection of the $Z\to b\bar b$ coupling, and for a strong sector invariant under a $U(2)^3$ flavour symmetry.
In the complete numerical analysis, all relevant constraints from $\Delta F=2$ processes are taken into account.
In all scenarios, visible effects in rare $K$ and $B$ decays like $K\to\pi\nu\bar\nu$, $B_s\to\mu^+\mu^-$ and $B\to K^*\mu^+\mu^-$ are possible that can be scrutinized experimentally in the near future.
Characteristic correlations between observables allow to distinguish the different cases.
To sample the large parameter space of the anarchic models, a new method is presented that allows larger statistics than conventional approaches.
}

\notoc
\begin{document}
\maketitle
\vspace{4cm}

\section{Introduction}

The discovery of a Higgs boson with Standard-Model-like properties faces us with a severe fine-tuning problem.
A straightforward solution to this puzzle would be a composite nature of the Higgs scalar.
To construct viable models with a composite Higgs boson, fermion masses have to be generated without at the same time violating the stringent bounds on flavour-changing neutral currents (FCNCs). A compelling mechanism to achieve this goal is partial compositeness, where the SM fermions are assumed to couple linearly to heavy, composite fermions with the same quantum numbers \cite{Kaplan:1991dc}.
Fermion masses are generated at low energies in a see-saw-like manner and the mass eigenstates are superpositions of elementary and composite fields, the degree of compositeness being larger for heavier fermions. This mechanism, explaining the fermion mass hierarchies by hierarchical composite-elementary mixings, in turn leads to a suppression of FCNCs even if the strong sector is completely flavour-anarchic \cite{Grossman:1999ra,Huber:2000ie,Gherghetta:2000qt}. Still, barring cancellations, the strong bounds from CP violation in kaon mixing require the resonances of the strong sector to be in excess of 10~TeV, in tension with naturalness \cite{Agashe:2004cp,Csaki:2008zd}. Consequently, various mechanisms have been suggested to further suppress flavour violation, one possibility being a flavour symmetry under which the strong sector is invariant, only broken by the composite-elementary mixings \cite{Cacciapaglia:2007fw,Barbieri:2008zt,Redi:2011zi,Barbieri:2012uh,Redi:2012uj}\footnote{Alternative solutions 
include flavour symmetries broken also in the strong sector \cite{Fitzpatrick:2007sa,Santiago:2008vq,Csaki:2008eh} or an extension of the (flavour-blind) global symmetry of the strong sector \cite{Bauer:2011ah}.}.

Another challenge of these models are electroweak precision tests. To avoid excessive contributions to the $T$ parameter, the strong sector should be invariant under the custodial symmetry $SU(2)_L\times SU(2)_R\times U(1)_X$, containing SM hypercharge as $Y=T_{3R}+X$ \cite{Agashe:2003zs}. Moreover, the couplings of the $Z$ boson to fermions, measured with high precision at LEP, are modified by two effects. First, since the heavy fermions come in vector-like pairs, after electroweak symmetry breaking (EWSB) the SM fermions mix with states that have different electroweak quantum numbers. Second, since the composite states are charged under the enlarged custodial symmetry, the mixing of vector resonances associated to $SU(2)_L$ and $SU(2)_R$ leads to additional corretions after EWSB.
Both effects can also lead to flavour-changing $Z$ couplings after going to the mass eigenstate basis. The resulting tree-level FCNCs put additional constraints on these models.
The strongest flavour-conserving constraint, which comes from the $Z$ coupling to left-handed $b$ quarks, can be avoided by making the strong sector (approximately) invariant under a discrete symmetry, which can be achieved by judicious choice of fermion representations \cite{Agashe:2006at}.
This choice will affect flavour-changing $Z$ couplings as well, leaving a characteristic pattern of effects than can be probed by FCNC measurements.

The experimental prospects for probing these flavour-changing $Z$ couplings in the near future are excellent.
In the $b\to s$ sector, the recent measurement of the decay $B_s\to\mu^+\mu^-$ \cite{:2012ct} has excluded sizable scalar current contributions to this mode and has started to become sensitive to $Z$-mediated effects. At the same time, the ongoing measurements of angular observables and CP asymmetries in the $B\to K^*\mu^+\mu^-$ decay will allow to disentangle chiralities and phases of the flavour-changing $Z$ couplings.
In the $s\to d$ sector, improved experiments probing the  decays $K_L\to\pi^0\nu\bar\nu$ and  $K^+\to\pi^+\nu\bar\nu$ are imminent.

The goal of this paper is a detailed numerical study of the effects of flavour-changing $Z$ couplings in models with partial compositeness. The focus will be on two different choices for the heavy fermion representations, both designed to avoid excessive contributions to the $Zb\bar b$ coupling, as well as on two different flavour structures: first, a scenario with an anarchic strong sector; second, a scenario with a strong sector invariant under a $U(2)^3$ flavour symmetry \cite{Barbieri:2011ci,Barbieri:2012uh,Barbieri:2012tu}, minimally broken by the composite-elementary mixings.
The basic structure of these models and the order of magnitude of the relevant effects have recently been analyzed in ref.\ \cite{Barbieri:2012tu}. Here, they will be scrutinized by a complete numerical study.
The main novelties of this paper are
\begin{itemize}
\item a first full numerical analysis of the custodially protected flavour-anarchic model where the left-handed up- and down-type quarks couple to two different composite fermions (to be called bidoublet model below),
\item a full numerical analysis of the custodially protected flavour-anarchic model where the left-handed quarks couple to a single composite fermion (to be called triplet model below), which has already been analyzed numerically in a similar framework in the literature \cite{Blanke:2008zb,Blanke:2008yr}, using updated flavour constraints (e.g.\ the $B_s$ mixing phase),
\item a novel method to sample the parameter space of partially composite models with flavour anarchy that allows much larger statistics than conventional methods,
\item a first full numerical analysis of the model with a $U(2)^3$ flavour symmetry in the strong sector \cite{Barbieri:2012uh,Barbieri:2012tu}, minimally broken by the composite-elementary mixings of right-handed quarks.
\end{itemize}

The paper is organized as follows. In section~\ref{sec:setup}, the definition  of the framework and the different scenarios will be recalled. In section~\ref{sec:analytic}, analytical expressions for the $Z$ couplings in the different scenarios will be derived to obtain a qualitative understanding of the effects in the numerical analysis. Section~\ref{sec:obs} will briefly review the relevant FCNC observables. Section~\ref{sec:numerics} is devoted to the numerical analysis. The summary and conclusions will be presented in section~\ref{sec:conc}. The novel method to scan the parameter space of anarchic models is discussed in appendix~\ref{sec:mcmc}.

\section{Setup}\label{sec:setup}

A simple effective framework to study the phenomenology of the models of interest is given by the two-site approach of ref.~\cite{Contino:2006nn}. In this framework, one considers one set of fermion resonances with heavy Dirac masses as well as spin-1 resonances associated to the global symmetry $SU(3)_c\times SU(2)_L\times SU(2)_R\times U(1)_X$ (``heavy gauge bosons''). This approach can be viewed as a truncation of 5D warped models, taking into account only the lightest set of KK states. This approximation is particularly justified in the case of tree-level amplitudes, which are the only ones relevant for this study. As a further simplification, all vector resonances will be assumed to have a common mass and coupling, $m_\rho$ and $g_\rho$, respectively.

As mentioned in the introduction, we will consider two different choices for the heavy fermion representations, dubbed triplet and bidoublet models.
Starting with the {\bf triplet model}\footnote{%
The triplet model was called TS10 (the two-site model with the fermions fitting into a \textbf{10} of $SO(5)\times U(1)_X$) in ref.~\cite{Vignaroli:2012si}.
A variation of the triplet model is obtained by having the $u_R$ mix with an additional $(1,1)_{2/3}$ instead of the $U$ from the triplet (see e.g.\  \cite{Blanke:2008zb,Blanke:2008yr,Albrecht:2009xr,Casagrande:2010si}).
While the protection of $Z$ couplings is the same as in the triplet model, due to the larger number of parameters, FCNCs effects tend to be even bigger. We will stick to the minimal case without the singlet in the following.
}, its fermion Lagrangian reads
\begin{gather}
\mathcal L_m^\text{} =\
-\text{tr}[ \bar L^i m_{L}^{ij} L^j ]
-\text{tr}[ \bar R^i m_{R}^{ij} R^j ]
-\text{tr}[ \bar R^{\prime\, i} m_{R}^{ij}R^{\prime\, j}]\notag\\
\mathcal L_Y^\text{} =\
Y^{ij} \text{tr}[ \bar L_L^i \mathcal H  R_R^j] + Y^{ij}\text{tr}[\mathcal H\,  \bar L_L^i R_R^{\prime\, j}]  + \text{h.c.}
\label{tripletL} \\
\mathcal L_\text{mix}^\text{} =
\lambda_{L}^{ij}\bar q_L^i Q_{R}^j
+
\lambda_{Ru}^{ij}\bar U_L^i u_{R}^j
+
\lambda_{Rd}^{ij}\bar D_L^i d_{R}^j
\end{gather}
where capital letters denote composite fields, which transform under $SU(2)_L\times SU(2)_R\times U(1)_X$ as
\begin{align}
L
=(Q ~ Q')
&=
\begin{pmatrix}
T & T_{\frac{5}{3}} \\
B & T_{\frac{2}{3}}
\end{pmatrix}
\sim(2,2)_\frac{2}{3}
\,,&
R&=
\begin{pmatrix}
U_{\frac{5}{3}} \\ U \\ D
\end{pmatrix}
\sim(1,3)_\frac{2}{3}
\,,&
R'&=
\begin{pmatrix}
U'_{\frac{5}{3}} \\ U' \\ D'
\end{pmatrix}
\sim(3,1)_\frac{2}{3}
\,.
\end{align}
While the strong sector $\mathcal L_m^\text{}+\mathcal L_Y^\text{}$ is invariant under the full custodial group, the composite-elementary mixings break $SU(2)_R\times U(1)_X$ explicitly.

The Lagrangian of the {\bf bidoublet model}\footnote{%
The bidoublet model was denoted TS5 in refs.~\cite{Bini:2011zb,Vignaroli:2012si} (the fermions fit into two \textbf{5} of $SO(5)\times U(1)_X$). 
}
reads
\begin{gather}\label{bidoubletL}
\mathcal L_m^\text{} =\
-\text{tr}[ \bar L_U^i m_{Q_u}^{ij} L_U^j ]
-\bar U^i m_{U}^{ij} U^j
+ (U\to D)
\\
\mathcal L_Y^\text{} =\
\left( Y_U^{ij} \text{tr}[ \bar L_U^i \mathcal H ]_L U_R^j  + \text{h.c.} \right)
+ (U\to D)
\\
\mathcal L_\text{mix}^\text{} =
\lambda_{Lu}^{ij}\bar q_L^i Q_{Ru}^j
+
\lambda_{Ru}^{ij}\bar U_L^i u_{R}^j
+ (U,u\to D,d)
\end{gather}
where
\begin{gather}
L_U=
(Q_u ~ Q_u')
=
\begin{pmatrix}
T & T_{\frac{5}{3}} \\
B & T_{\frac{2}{3}}
\end{pmatrix}
\sim(2,2)_\frac{2}{3}
\,,~
L_D=
(Q_d' ~ Q_d)
=
\begin{pmatrix}
B_{-\frac{1}{3}} & T' \\
B_{-\frac{4}{3}} & B'
\end{pmatrix}
\sim(2,2)_\frac{1}{3}
\,,
\\
U \sim(1,1)_\frac{2}{3}
\,,\qquad
D \sim(1,1)_{-\frac{1}{3}}
\,.
\label{eq:BD-fields}
\end{gather}
An important new feature with respect to the triplet model is that the left-handed up- and down-type quarks now couple to two different composite fields, so that the degrees of compositeness of the left-handed top and bottom quarks, for instance, are unrelated. This is crucial not only for the suppression of the corrections to the $Z\to b\bar b$ coupling, as discussed in section~\ref{sec:analytic}, but is also necessary in models with a flavour symmetry only broken by the left-handed composite-elementary mixings \cite{Redi:2011zi}.

For the flavour structure, we consider two possibilities: anarchy and $U(2)^3$. In the anarchic case, the strong sector has no flavour symmetry and the quark mass and mixing hierarchies are generated by hierarchies in the composite-elementary mixings.
In the bidoublet model, the Yukawa couplings $Y_{U,D}^{ij}$ are anarchic matrices, while the composite-elementary mixings $\lambda_X^{ij}$ can be chosen to be real and diagonal.
In the triplet model, a basis where $\lambda_X^{ij}$ are real and diagonal can be chosen as well, although $SU(2)_R$ is no longer manifest in this basis and one effectively gets different Yukawas in the up- and down-type sectors, related by $Y_{U}^{ij}=Y_{D}^{ik} U^{kj}$, where $U$ is a unitary matrix.
For simplicity, the Dirac mass terms will be assumed to be proportional to the identity in the following\footnote{Relaxing this assumption will not lead to qualitatively new effects, but might allow larger effects in view of the enlarged parameter space.}.

In the $U(2)^3$ case, the strong sector is assumed to be invariant under a flavour symmetry under which the first two generation fields transform as doublets and the third generation as singlets \cite{Barbieri:2011ci}. The symmetry is broken by the left-handed {\em or} right-handed composite-elementary mixings \cite{Barbieri:2012uh,Barbieri:2012tu}.
The breaking is assumed to be minimal, i.e.\ using the smallest set of small symmetry breaking spurions that is required to generate the observed quark masses and mixings.
The case where the breaking occurs in the right-handed mixings is called left-handed compositeness. In this case, the Dirac mass matrices, Yukawa matrices and left-handed mixings are all of the form $\text{diag}(a,a,b)$ with real entries. The form of the right-handed mixings compatible with the minimal breaking of $U(2)^3$ can be found in the appendix of ref.~\cite{Barbieri:2012tu}. Analogously, in right-handed compositeness, the right-handed mixings are diagonal while the left-handed ones contain the $U(2)^3$ breaking spurions.

A further attractive possibility is that the strong sector is fully flavour-invariant, i.e.\ possesses a $U(3)^3$ symmetry \cite{Cacciapaglia:2007fw,Barbieri:2008zt,Redi:2011zi}. In this case however, flavour-changing $Z$ couplings do not arise at tree-level at all \cite{Barbieri:2012tu}, so it will not be considered in the following.

\section{\textit{Z} couplings}\label{sec:analytic}

In ref.~\cite{Agashe:2006at} it has been shown that two discrete subgroups of the custodial group can protect the $Zb\bar b$ coupling from excessive tree-level corrections, as required by electroweak precision measurements: the $P_{LR}$ symmetry, which exchanges $SU(2)_L$ and $SU(2)_R$, and the $P_C$ symmetry, under which $T_{3L,R}\to-T_{3L,R}$. In the triplet model, the $d_L^i$ mix with the $P_{LR}$ eigenstates $B^i$, while the $u_R^i$ mix with the $P_C$ eigenstates $U^i$. Consequently, the (flavour-conserving and flavour-changing \cite{Blanke:2008zb,Buras:2009ka}) $Z$ couplings to left-handed down-type and right-handed up-type quarks at zero momentum receive no correction at tree-level.
In the bidoublet model, the right-handed quarks mix with the $P_C$ and $P_{LR}$ eigenstates $U$ and $D$, so the right-handed $Z$ couplings are not corrected. In the case of left-handed mixings, the situation is slightly more involved compared to the triplet model, since the left-handed elementary doublet now mixes with two different fields, $Q_u$ and $Q_d$, 
via the mixing 
terms $\lambda_{Lu}$ and $\lambda_{Ld}$. While the $T'$ and $B$ (in the notation of eq.~(\ref{eq:BD-fields})) are $P_{LR}$ eigenstates, the $T$ and $B'$ are not. Consequently, the $Zd^i_L\bar d^j_L$ couplings receive corrections involving $\lambda_{Ld}$, but not $\lambda_{Lu}$. This is enough to protect the $Zb_L\bar b_L$ coupling, since a hierarchy $\lambda_{Ld}\ll \lambda_{Lu}$ is natural in view of the lightness of the $b$ quark compared to the top quark.

\begin{table}[tbp]
\renewcommand{\arraystretch}{1.3}
\centering
\begin{tabular}{ccccc}
\hline
& $\delta g_{Ld}$& $\delta g_{Rd}$& $\delta g_{Lu}$&$\delta g_{Ru}$\\
\hline
triplet & 0 & $- (\tfrac{1}{2}\Delta_{Rd} + \Delta_g) s_{Rd}^2$ & $- (\Delta_{Lu} + \Delta_g) s_{L}^2$ & 0 \\
bidoublet & $(\tfrac{1}{2} \Delta_{Ld} + \Delta_g) s_{Ld}^2$ & 0 & $- (\tfrac{1}{2}\Delta_{Lu} + \Delta_g) s_{Lu}^2$ & 0 \\
\hline
\end{tabular}
\caption{Tree-level contributions to the modified $Z$ couplings as defined in eq.~(\ref{eq:zcoup}) in the triplet and bidoublet models. The $\Delta_{X}$ are defined in eq.~(\ref{eq:delta}).}
\label{tab:zcoup}
\end{table}

To obtain approximate analytical expressions for the modified $Z$ couplings, let us start with a simplified one-generation model.
Writing the $Z$ couplings to quarks as
\begin{equation}
\frac{g}{c_w}\bar q \gamma^\mu \left[
(T_{3L}-Qs_w^2+\delta g_{Lq}^{}) P_L
+(-Qs^2_w+\delta g_{Rq}^{})P_R
\right]q
\,Z_\mu
\,,
\label{eq:zcoup}
\end{equation}
the contributions due to fermion and gauge boson mixing can then be written as in table~\ref{tab:zcoup}, where
\begin{equation}
s_{X}={\lambda_{X}}/{\sqrt{m_{X}^2+\lambda_{X}^2}}
\end{equation}
is the {\em degree of compositeness} and
\begin{align}
\Delta_{Lu} &= \frac{v^2Y_{U}^2}{2m_U^2}
~,&
\Delta_{Ld} &= \frac{v^2Y_{D}^2}{2m_D^2}
~,&
\Delta_{Rd} &= \frac{v^2Y_{D}^2}{2m_{L}^2}
~,&
\Delta_{g} &= \frac{v^2g_\rho^2}{4m_\rho^2}
~.
\label{eq:delta}
\end{align}
In the three-generation case, (\ref{eq:delta}) remains valid for the flavour-diagonal couplings\footnote{In anarchy, this is true if one interprets the $Y_{U,D}$ as ``average'' anarchic Yukawas.}, in particular for the $Z\to b\bar b$ coupling, while the flavour-violating couplings are dependent on the flavour structure.
In anarchy, all the flavour-changing couplings allowed by the $P_{LR}$ and $P_C$ symmetries are generated, while in $U(2)^3$, the right-handed flavour-changing couplings are strongly suppressed. In fact, in $U(2)^3$ with right-handed compositeness, no flavour-changing couplings are generated at all at tree-level \cite{Barbieri:2012tu}. A generic prediction of minimally broken $U(2)^3$ is that the amplitudes of 12-transitions are aligned in phase with the SM \cite{Barbieri:2011ci}.

With these considerations, one can already recognize a pattern among the flavour-changing $Z$ couplings allowed by the different electroweak and flavour structures that can potentially serve to distinguish them experimentally. Table~\ref{tab:dna} lists the presence or absence of the flavour-changing couplings relevant for $K$, $B_{d,s}$ and $D$ decays for the models discussed so far.

\begin{table}[tbp]
\centering
\renewcommand{\arraystretch}{1.3}
\begin{tabular}{cccccccc}
\hline
&& \multicolumn{2}{c}{$K$} & \multicolumn{2}{c}{$B_{d,s}$} & \multicolumn{2}{c}{$D$} \\
&& $L$ & $R$ & $L$ & $R$ & $L$ & $R$ \\
\hline
\multirow{2}{*}{\CircledA}
& triplet &  & $\mathbbm{C}$ &  & $\mathbbm{C}$ & $\mathbbm{C}$ &  \\
& bidoublet & $\mathbbm{C}$ &  & $\mathbbm{C}$ &  & $\mathbbm{C}$ &  \\
\multirow{2}{*}{$U(2)^3_\text{LC}$}
& triplet &  &  &  &  & $\mathbbm{R}$ &  \\
& bidoublet & $\mathbbm{R}$ &  & $\mathbbm{C}$ &  & $\mathbbm{R}$ &  \\
\hline
\end{tabular}
\caption{Pattern of effects in flavour-changing $Z$ couplings relevant for $K$ ($s\to d$), $B$ ($b\to s,d$) and $D$ ($c\to u$) physics.
$\mathbbm{R}$ denotes contributions aligned in phase with the SM, $\mathbbm{C}$ contributions with a new CPV phase and empty cells refer to vanishing or strongly suppressed contributions.}
\label{tab:dna}
\end{table}

One can now go one step further and estimate the allowed size of the flavour-changing couplings in the various models, to understand the size of the effects encountered in the full-fledged numerical analysis of section~\ref{sec:numerics}.
To this end, it turns out to be convenient to consider rescaled couplings
\begin{equation}
\delta_{L,R}^{ij} = \frac{8\pi^2 s_w^2}{e^2\xi_{ij}} \,\delta g_{Ld,Rd}^{ij} ~,
\label{eq:deltaLR}
\end{equation}
where $\xi_{ij} = V_{tj}V_{ti}^*$. This definition is chosen so
that $O(1)$ values of the $\delta_{L,R}^{ij}$ correspond roughly to the size of the loop-induced flavour-changing $Z$ coupling in the SM (and thus can lead to visible effects in FCNC observables, see section~\ref{sec:obs}) and for real $\delta_{L,R}^{ij}$, the couplings are aligned with the SM contribution.

\paragraph{Anarchic triplet model} Here the flavour-changing couplings can be estimated by replacing $s_R^2 \to s_R^is_R^j$ in table~\ref{tab:zcoup}. Using the well-known approximate relations\footnote{replacing all functions of Yukawa couplings by appropriate powers of an ``average'' Yukawa $Y_*$} between the degrees of compositeness, the CKM angles and the quark masses,
\begin{align}
y_{d^i} &\sim Y_* s_{Ld}^is_{Rd}^i
\,,&
V_{ij} &\sim s_{Ld}^i/s_{Ld}^j
\,,&
\end{align}
and defining $x_t=s_{Lt}/s_{Rt}\sim O(1)$,
one can write the right-handed flavour-changing couplings to down-type quarks as 
\begin{align}
\delta g_{Rd}^{ij} &\sim \frac{y_{d^i}y_{d^j}}{\xi_{ij}x_t y_t Y_* z^2} (\tfrac{1}{2}\Delta_{Rd} + \Delta_g)
\,,
\end{align}
leading to
\begin{align}
|\delta_R^{12,13,23}| &\sim 
\lbrace
0.3,0.02,0.02
\rbrace
\left[\frac{(\tfrac{1}{2}\Delta_{Rd} + \Delta_g)}{0.1}\right]
\left[\frac{1}{x_t}\right]
\left[\frac{3}{Y_*}\right]
\,.
\end{align}
Consequently, one can expect sizable effects in $K$ decays, while effects in $B$ or $B_s$ decays should be smaller.
Note that the size of these effects is not constrained by $Z\to b\bar b$, which mostly constrains the {\em left-handed} $Z\bar bb$ coupling (cf.\ \cite{Barbieri:2012tu,Guadagnoli:2013mru}).

\paragraph{Anarchic bidoublet model} Replacing accordingly $s_L^2 \to s_L^is_L^j$ in table~\ref{tab:zcoup}, one can estimate the left-handed flavour-changing couplings in terms of the coupling to $b$ quarks as
\begin{align}
\delta g_{Ld}^{ij} &\sim \xi_{ij} ~ \delta g_{Ld}^{33}~.
\end{align}
Since $\delta g_{Ld}^{33}$ is constrained by $Z\to b\bar b$ to be less than about $5\times10^{-4}$, one finds
\begin{align}
|\delta_L^{12,13,23}| &\sim
0.1
\left[\frac{\delta g_{Ld}^{33}}{5\times10^{-4}}\right]
\,.
\end{align}
Consequently, one expects effects in $K$ physics to be smaller than in the triplet model, but could have sizable effects in $B$ physics.

\paragraph{\boldmath $U(2)^3$ bidoublet model}
In this case, the left-handed flavour-changing couplings to down-type quarks can be estimated as
\begin{align}
\delta g_{Ld}^{i3} &\approx  \xi_{i3} \, r_b \, \delta g_{Ld}^{33}~,
&
\delta g_{Ld}^{12} &\approx  \xi_{12} \, |r_b|^2 \, \delta g_{Ld}^{33}~,
\label{eq:ZbsL-U2}
\end{align}
where $r_b$ is a complex parameter with magnitude of $O(1)$.
Similarly to the anarchic bidoublet model above, one thus obtains
\begin{align}
|\delta_L^{12,13,23}| &\sim
0.1
\,
\lbrace
|r_b|^2,|r_b|,|r_b|
\rbrace
\,
\left[\frac{\delta g_{Ld}^{33}}{5\times10^{-4}}\right]
\,.
\end{align}
For $|r_b|\sim 1$, one could thus have visible effects in both $K$ and $B$ decays, while for $|r_b|<1$ ($>1$) effects in $B$ ($K$) physics should be larger.

\section{Probes of flavour-changing \textit{Z} couplings}\label{sec:obs}

The flavour-changing $Z$ couplings contribute to FCNC processes mediated by $Z$ exchange in the SM, i.e.\ rare $K$, $B$, $B_s$ and $D$ decays. Below some formulae for the relevant observables are collected. 

\subsection{\texorpdfstring{\boldmath $s\to d$}{s->d}}\label{sec:sd}

The branching ratios of the charged and neutral $K\to \pi\nu\bar\nu$ decays can be written as
\begin{align}
\label{bkpnZ}
{\rm BR}(K^+\to \pi^+\nu\bar\nu) &= \kappa_+(1+\Delta_\text{EM})\left[\left(\frac{{\rm Im}(\xi_{12} X_K)}{\lambda^5}\right)^2 
+ \left( - P_{(u,c)} + \frac{{\rm Re}(\xi_{12} X_K)}{\lambda^5}\right)^2\right]
\\
\label{bklpnZ}
{\rm BR}(K_L\to \pi^0\nu\bar\nu ) &= \kappa_L \left( \frac{{\rm Im}(\xi_{12} X_K)}{\lambda^5}\right)^2
\end{align}
where
$X_K=X_\text{SM}+(\delta_L^{12} + \delta_R^{12})$ with $X_\text{SM}=1.469 \pm 0.017$ \cite{Brod:2010hi}. The quantities $\kappa_+$, $\kappa_L$ and $P_{(u,c)}$ can be found e.g.\ in \cite{Brod:2010hi}. The experimental measurement and SM predicitions read \cite{Brod:2010hi}
\begin{align}
\text{BR}(K^+\to \pi^+\nu\bar\nu)_\text{exp} &=  (17.3^{+11.5}_{-10.5})\times10^{-11}
\,,\\
\text{BR}(K^+\to \pi^+\nu\bar\nu)_\text{SM} &= (7.81\pm0.80)\times10^{-11}
\,,\\
\text{BR}(K_L\to \pi^0\nu\bar\nu)_\text{SM} &=  (2.43\pm0.39)\times10^{-11}
\,,
\end{align}
while the experimental bound on $K_L\to \pi^0\nu\bar\nu$ has not yet reached the model-independent theoretical upper bound \cite{Grossman:1997sk}.

Another relevant process is the decay $K_L\to \mu^+\mu^-$. Unfortunately, it is plagued by long-distance contributions which are hard to calculate. The short-distance part can be written as
\begin{align}
\label{bkmm}
{\rm BR}(K_L\to \mu^+\mu^-)_\text{SD} &= \kappa_\mu\left( - \bar P_{c} + \frac{{\rm Re}(\xi_{12} Y_K)}{\lambda^5}\right)^2,
\end{align}
where
$Y_K=Y_\text{SM}+(\delta_L^{12} - \delta_R^{12})$ with $Y_\text{SM}=0.98\pm0.02$ and the quantity $\bar P_{c}$ can be found in \cite{Buchalla:1995vs}. An estimate of the long-distance contributions allows to extract an experimental upper bound on the short-distance part~\cite{Isidori:2003ts}
\begin{equation}
{\rm BR}(K_L\to \mu^+\mu^-)_\text{SD} < 2.5\times 10^{-9}\,.
\label{eq:Kmmbound}
\end{equation}
Since $K_L\to \mu^+\mu^-$ depends on the difference of left- and right-handed couplings, while the $K\to \pi\nu\bar\nu$ decays depend on their sum, the constraints are complementary.

\subsection{\texorpdfstring{\boldmath $b\to s$}{b->s}}

The $b\to s$ transition is the one best constrained experimentally due to a multitude of inclusive and exclusive decays sensitive to it. For our purposes, the most relevant decays are $B_s\to\mu^+\mu^-$ and $B\to K^*\mu^+\mu^-$.

In the absence of scalar currents, the branching ratio of the $B_s\to\mu^+\mu^-$ decay can be written as
\begin{equation}
\frac{\text{BR}(B_s\to\mu^+\mu^-)}{\text{BR}(B_s\to\mu^+\mu^-)_\text{SM}}
=\left| 1+\frac{\delta_L^{23}-\delta_R^{23}}{Y_\text{SM}} \right|^2,
\label{eq:Bsmumu}
\end{equation}
where $Y_\text{SM}$ was given after eq.~(\ref{bkmm}).
For the flavour-averaged branching ratio, one currently has \cite{DeBruyn:2012wk,Buras:2012ru,:2012ct}
\begin{align}
\text{BR}(B_s\to\mu^+\mu^-)_\text{exp} &= (2.9^{+1.4}_{-1.1})\times 10^{-9} \,,
\\
\text{BR}(B_s\to\mu^+\mu^-)_\text{SM} &= (3.23\pm0.27)\times 10^{-9} \,.
\end{align}
Eq.~(\ref{eq:Bsmumu}) is also valid for the decay $B_d\to\mu^+\mu^-$ with the replacement $\delta_X^{23}\to \delta_X^{13}$.

The $B\to K^*\mu^+\mu^-$ decay allows access to many observables sensitive to new physics by means of an angular analysis. Since only modified $Z$ couplings are of interest here, only the region of large dilepton invariant mass, $q^2>14.18\,\text{GeV}^2$, will be considered. This region has two advantages. First, magnetic dipole operators, which can also be generated in the models at hand but are not taken into account here, do not contribute in this region. Second, this region is theoretically cleaner since non-factorizable effects are under better control \cite{Beylich:2011aq} and since the form factors can be calculated by lattice QCD.

While the precise expressions for the $B\to K^*\mu^+\mu^-$ observables can be found e.g.\ in ref.~\cite{Altmannshofer:2008dz}, 
below approximate numerical expressions for the central values of theory parameters are reported
to understand the sensitivity of the observables to the $Z$ couplings (in the numerical analysis the exact formulae are used).
For the branching ratio, one finds
\begin{multline}
\frac{\text{BR}(B\to K^*\mu^+\mu^-)}{\text{BR}(B\to K^*\mu^+\mu^-)_\text{SM}}
\approx
1
+1.3 \,\text{Re}(\delta_L^{23})
-0.91 \,\text{Re}(\delta_R^{23})
\\
+0.63 \,\left(|\delta_L^{23}|^2+|\delta_R^{23}|^2\right)
-0.91 \,\text{Re}(\delta_L^{23}\delta_R^{23*})
\,.
\label{eq:BKsmumu}
\end{multline}
Among the angular observables, the forward-backward asymmetry $A_\text{FB}$, the angular CP asymmetry $A_9$ and the CP-averaged angular observable $S_3$ are particularly interesting, since they can be extracted from a one-dimensional angular distribution. Expanding their precise expressions to first order in the couplings $\delta_{L,R}^{23}$, one obtains for them
\begin{align}
\frac{A_\text{FB}(B\to K^*\mu^+\mu^-)}{A_\text{FB}(B\to K^*\mu^+\mu^-)_\text{SM}}
&\approx
1-0.12\,\text{Re}(\delta_L^{23})-0.93\,\text{Re}(\delta_R^{23})\,,
\label{eq:AFB}
\\
\frac{S_3(B\to K^*\mu^+\mu^-)}{S_3(B\to K^*\mu^+\mu^-)_\text{SM}}
&\approx
1-1.4\,\text{Re}(\delta_R^{23})\,,
\\
A_9(B\to K^*\mu^+\mu^-)
&\approx
-0.31\,\text{Im}(\delta_R^{23})\,,
\end{align}
valid for the high $q^2$ region, $q^2>14.18\,\text{GeV}^2$.
Importantly, the observables $S_3$ and $A_9$ are only sensitive to the right-handed coupling. Experimental averages and SM predicitions for all the observables can be found e.g.\ in ref.~\cite{Altmannshofer:2012az}.

While not accessible at the LHC, future $B$ factories will look for the decays $B\to K^{(*)}\nu\bar\nu$ (and possibly also the inclusive decay). Being free from photon penguin contributions, also these decays are clean probes of modified $Z$ couplings.
Defining
\begin{equation}  \label{eq:epsetadef}
 \epsilon = \frac{\sqrt{ |\delta^{23}_L|^2 + |\delta^{23}_R|^2}}{X_\text{SM}}~, \qquad
 \eta = \frac{-\text{Re}\left(\delta^{23}_L \delta^{23*}_R\right)}{|\delta^{23}_L|^2 + |\delta^{23}_R|^2}~,
\end{equation}
where $X_\text{SM}$ was given in section~\ref{sec:sd}, the branching ratios and the angular observable $F_L$ in $B\to K^*\nu\bar\nu$ can be written as \cite{Altmannshofer:2009ma}
\begin{align}
\frac{\text{BR}(B\to K\nu\bar\nu)}{\text{BR}(B\to K\nu\bar\nu)_\text{SM}} & =
 (1 - 2\,\eta)\epsilon^2~, &
\frac{\text{BR}(B\to K^*\nu\bar\nu)}{\text{BR}(B\to K^*\nu\bar\nu)_\text{SM}} & =
(1 + 1.31 \,\eta)\epsilon^2~, \\
\frac{\text{BR}(B\to X_s\nu\bar\nu)}{\text{BR}(B\to X_s\nu\bar\nu)_\text{SM}} & =
(1 + 0.09 \,\eta)\epsilon^2~, &
\frac{F_L(B\to K^*\nu\bar\nu)}{F_L(B\to K^*\nu\bar\nu)_\text{SM}} & =
\frac{(1 + 2 \,\eta)}{(1 + 1.31 \,\eta)}~.
\end{align}
A combined analysis of these observables would thus allow to disentangle left- and right-handed couplings.

\section{Numerical analysis}\label{sec:numerics}

After estimating the size of the flavour-changing couplings in section~\ref{sec:analytic} and defining the relevant experimental observables in section~\ref{sec:obs}, we are now ready for a numerical analysis of the effects in the various models considered. Such analysis is particularly challenging in the anarchic case in view of the large number of free parameters and the necessity to reproduce the known quark masses and mixings. To this end, a novel approach was used in this study, based on a Markov chain Monte Carlo and described in more detail in appendix~\ref{sec:mcmc}. Here it suffices to say that the result corresponds to a sample of the parameter space with certain ranges allowed for the free parameters. Throughout the numerical analysis, a common mass and coupling, $m_\rho$ and $g_\rho$, was assumed for the vector resonances. In the anarchic case, as mentioned above, the further simplification of assuming the heavy fermion Dirac mass matrices to be proportional to the identity was made. The 
parameters were 
then allowed to float in the ranges
\begin{align}
g_\rho &\in [3,5] \,,
&
Y_{U,D}^{ij} &\in [0,4\pi] \,,
&
\text{arg}(Y_{U,D}^{ij}) &\in [0,2\pi] \,,
\\
m_\rho &\in [3,5] \,\text{TeV} \,,
&
m_\psi &\in [0.5,3] \,\text{TeV} \,,
\end{align}
where $m_\psi$ stands for any heavy fermion mass, with all the additional parameters unconstrained.
Since the goal of the present analysis is to obtain the maximum allowed deviations of the flavour-changing $Z$ couplings from the SM, the parameter ranges are chosen to allow for sizable effects. The lower bound on $m_\rho$ is chosen to fulfill the constraint from the $S$ parameter, while the upper bound on $g_\rho$ is chosen to always have $f=m_\rho/g_\rho>600$~GeV.
Concerning the range for the $Y_{U,D}^{ij}$, a magnitude of $4\pi$ is the largest value for which the effective theory description could be valid roughly until the mass of the first resonances \cite{Casagrande:2008hr}. Imposing a smaller maximum value $Y_\text{max}$ for the Yukawa couplings would reduce the maximally allowed effects in the numerical analysis roughly by a factor of $4\pi/Y_\text{max}$.
Note that, contrary to ref.~\cite{Barbieri:2012tu}, the $Y_{U,D}$ and the composite fermion masses are treated as unrelated.

Throughout the analysis, the relevant constraints from $\Delta F=2$ processes, namely the mass differences and mixing phases of the $B_d$ and $B_s$ systems and the CP violating parameter $|\epsilon_K|$, have been required to be within their experimentally allowed bounds (for details of the procedure see appendix~\ref{sec:mcmc}). As is well known, the $|\epsilon_K|$ constraint is particularly hard to fulfill in the flavour anarchic case. The approach  of this study is to accept the implied fine-tuning in the parameter space and to study the maximum allowed size of flavour-changing $Z$ couplings. In the $U(2)^3$ case, this problem is absent.

\subsection{Flavour-changing \textit{Z} couplings}

\begin{figure}[tbp]
\centering
\includegraphics[width=0.44\textwidth]{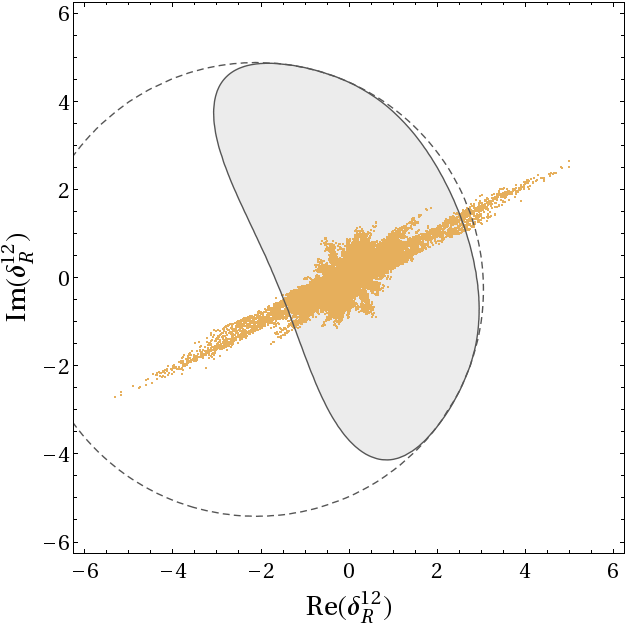}
\hspace{0.05\textwidth}%
\includegraphics[width=0.45\textwidth]{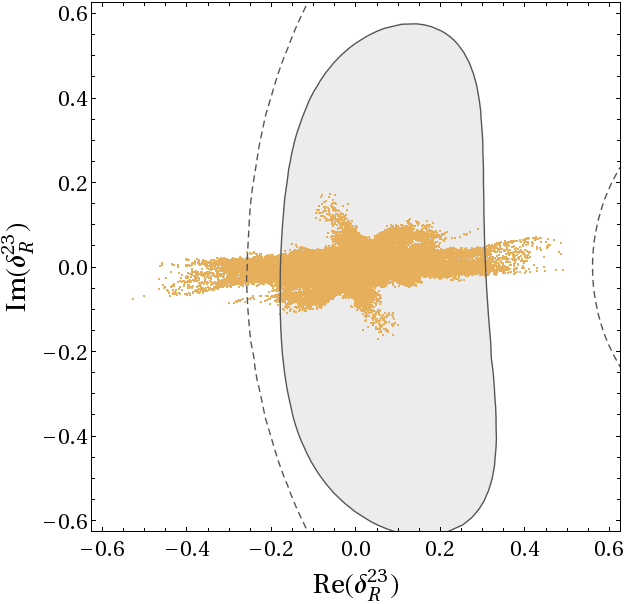}
\caption{Right-handed flavour-changing $Z$ couplings in the {\bf anarchic triplet model} relevant for $s\to d$ (left) and $b\to s$ transitions (right), normalized as in eq.~(\ref{eq:deltaLR}).
The gray regions show the combined experimental constraints at 95\%~C.L. In the left-hand plot, the dashed circle shows the constraint from $\text{BR}(K^+\to\pi^+\nu\bar\nu)$ alone. In the right-hand plot, the dashed ring shows the constraint from $\text{BR}(B_s\to\mu^+\mu^-)$ alone.}
\label{fig:zcoup_triplet}
\end{figure}

We start by looking at the size of $s\to d$ and $b\to s$ couplings in the various models. Figure~\ref{fig:zcoup_triplet} shows $\delta_R^{12}$ and $\delta_R^{23}$ in the anarchic triplet model, compared to the experimentally allowed ranges. All points satisfy the constraints from $\Delta F=2$ observables and $Z\to b\bar b$.
In the left-hand plot, the gray region shows the combined constraint from $\text{BR}(K^+\to\pi^+\nu\bar\nu)$ and $\text{BR}(K_L\to\mu^+\mu^-)_\text{SD}$ (treating eq.~(\ref{eq:Kmmbound}) as a 68\%~C.L.\ bound), while the dashed circle shows the constraint from $\text{BR}(K^+\to\pi^+\nu\bar\nu)$ alone. In the right-hand plot, the gray region shows the combined constraint from all relevant $B\to K^*\mu^+\mu^-$ observables at high $q^2$ and from $\text{BR}(B_s\to\mu^+\mu^-)$, while the dashed ring shows the constraint from $\text{BR}(
B_s\to\mu^+\mu^-)$ alone\footnote{For details on how 
these constraints are obtained numerically, the reader is referred to ref.~\cite{Altmannshofer:2012az}.}.
As expected from the estimates 
in section~\ref{sec:analytic}, the effects in $K$ physics are sizable, while the effects in $B$ physics are smaller (but still sizable compared to the more stringent experimental bounds).
Note that the tilted main branch in the left-hand plot corresponds to almost real $\delta g_{Rd}^{12}$, which is preferred due to the interplay between $\Delta S=1$ and $\Delta S=2$ amplitudes and the strong constraint on $|\epsilon_K|$. The tilt is caused by dividing out the complex SM CKM factors in eq.~(\ref{eq:deltaLR}).

\begin{figure}[tbp]
\centering
\includegraphics[width=0.44\textwidth]{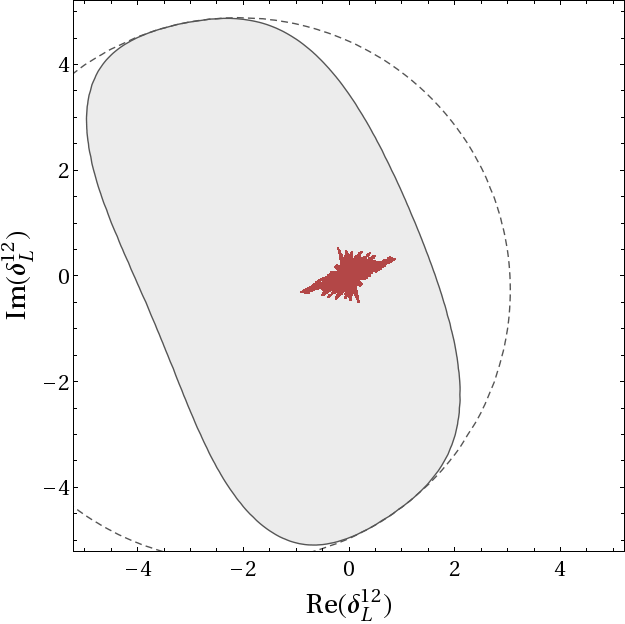}
\hspace{0.05\textwidth}%
\includegraphics[width=0.45\textwidth]{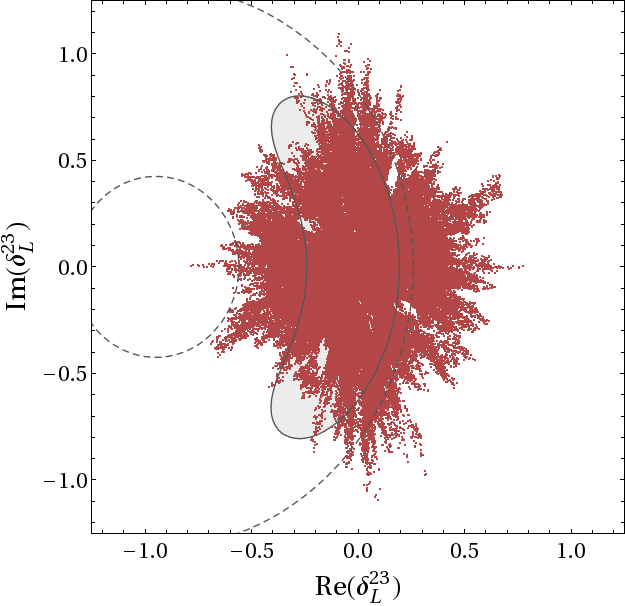}
\caption{Left-handed flavour-changing $Z$ couplings in the {\bf anarchic bidoublet model} relevant for $s\to d$ (left) and $b\to s$ transitions (right), normalized as in eq.~(\ref{eq:deltaLR}).}
\label{fig:zcoup_bidoublet}
\end{figure}

Figure~\ref{fig:zcoup_bidoublet} shows analogous plots in the anarchic bidoublet model, this time showing the {\em left-handed} couplings. Also here, the estimates of section~\ref{sec:analytic} are confirmed: while the effects in the $s\to d$ coupling are smaller than in the triplet case, the effects in $b\to s$ can be sizable. In fact, both $B\to K^*\mu^+\mu^-$ and $B_s\to\mu^+\mu^-$ already cut into the model's parameter space.

\begin{figure}[tbp]
\centering
\includegraphics[width=0.45\textwidth]{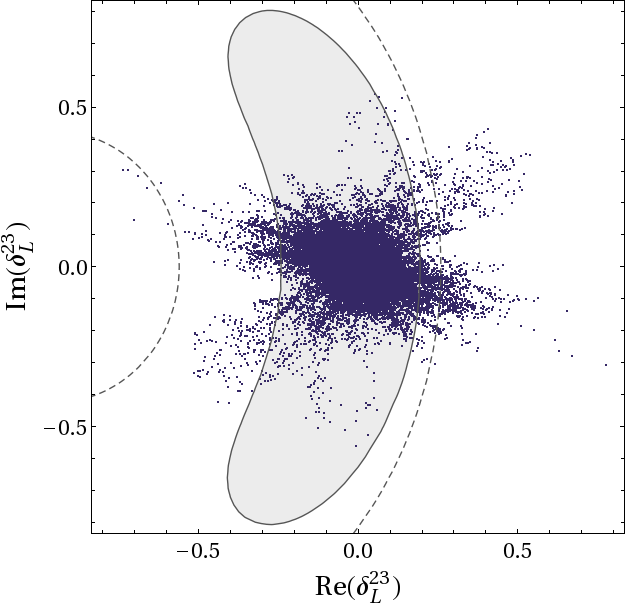}
\caption{Left-handed flavour-changing $Z$ couplings in the {\bf\boldmath bidoublet model with $U(2)^3$ left-handed compositeness} relevant for $b\to s$ transitions, normalized as in eq.~(\ref{eq:deltaLR}).}
\label{fig:zcoup_U2}
\end{figure}

Figure~\ref{fig:zcoup_U2} shows the size of the left-handed $b\to s$ couplings in the bidoublet model with $U(2)^3$ (left-handed compositeness). Just as in the anarchic case, one can obtain large effects in $B$ physics with a new CP violating phase. The main difference is that the size of the imaginary part is now limited. The reason is that, for the parameter $r_b$ in (\ref{eq:ZbsL-U2}) to deviate from 1 (in which case the $Z$ coupling would be aligned in phase with the SM) requires a relatively large degree of compositeness for the first two generations of left-handed up-type quarks, leading to a sizable correction to the hadronic width $R_h$ (see the discussion in the appendix of ref.~\cite{Barbieri:2012tu}).
The effects in the $s\to d$ couplings are similar in magnitude to the anarchic case, but now $\text{Im}(\delta_L^{12})=0$, which is why the corresponding plot is omitted.

\subsection{Predictions for rare decays}

Having established the size of the modified $Z$ couplings, we can now proceed to analyze their impact on the FCNC observables\footnote{%
Additional contributions to $\Delta F=1$ amplitudes from the exchange of neutral vector resonances are subleading and thus neglected. For $\Delta F=2$ constraints, they are instead important and taken into account.
}.

\begin{figure}[p]
\centering
\includegraphics[height=6.5cm]{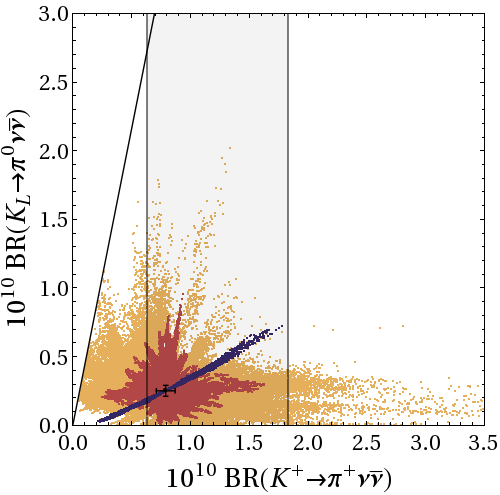}%
\hspace{0.05\textwidth}%
\includegraphics[height=6.5cm]{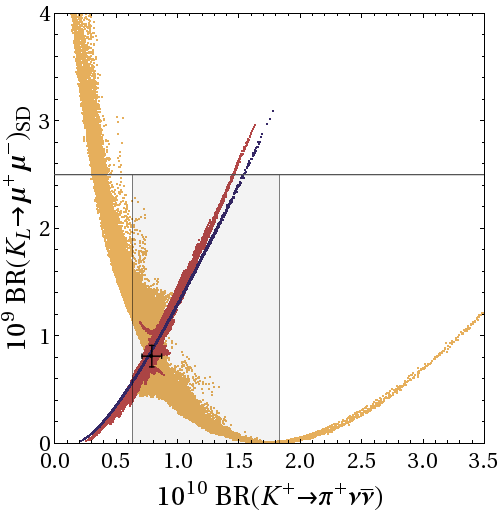}

\centering
\includegraphics[height=6.5cm]{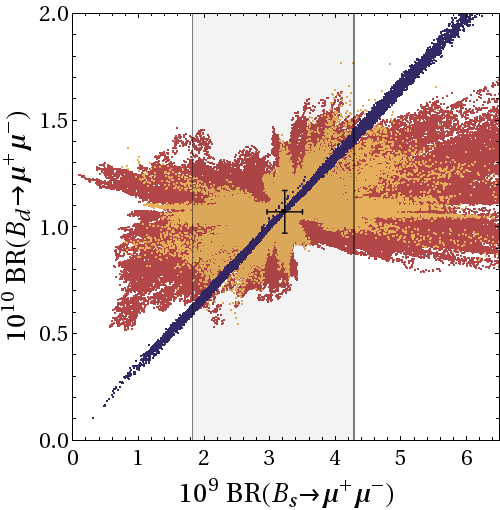}%
\hspace{0.05\textwidth}%
\includegraphics[height=6.5cm]{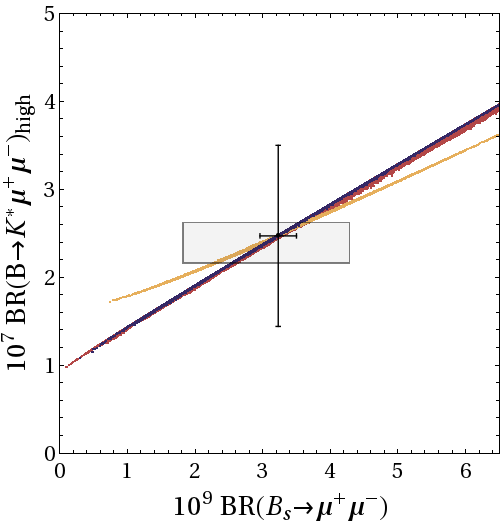}

\centering
\includegraphics[height=6.7cm]{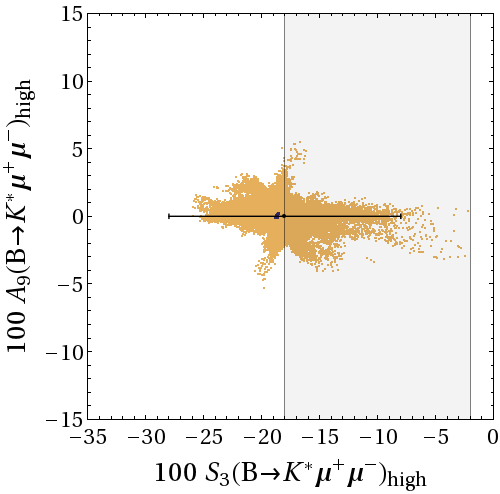}%
\hspace{0.05\textwidth}%
\includegraphics[height=6.5cm]{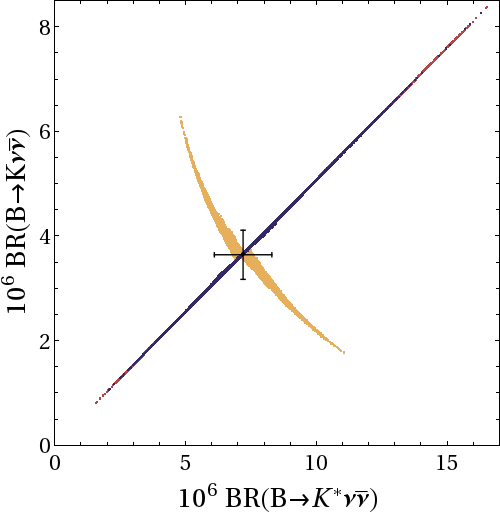}
\caption{Correlations between observables in rare $K$ and $B$ decays in the anarchic triplet model (yellow/light gray), the anarchic bidoublet model (red/medium gray) and the bidoublet model with $U(2)^3$ left-handed compositeness (blue/dark gray). The shaded regions indicate experimental $1\sigma$ ranges. The black cross shows the SM predictions with theory uncertainties.}
\label{fig:BR_all}
\end{figure}

Figure~\ref{fig:BR_all} shows the correlations between
$\text{BR}(K^+\to\pi^+\nu\bar\nu)$ and $\text{BR}(K_L\to\pi^0\nu\bar\nu)$,
between
$\text{BR}(K^+\to\pi^+\nu\bar\nu)$ and $\text{BR}(K_L\to\mu^+\mu^-)_\text{SD}$,
between
$\text{BR}(B_s\to\mu^+\mu^-)$ and $\text{BR}(B_d\to\mu^+\mu^-)$,
between
$\text{BR}(B_s\to\mu^+\mu^-)$ and $\text{BR}(B\to K^*\mu^+\mu^-)$ (in the high $q^2$ region),
between the $B\to K^*\mu^+\mu^-$ observables $S_3$ and $A_9$
as well as between
$\text{BR}(B\to K^*\nu\bar\nu)$ and $\text{BR}(B\to K\nu\bar\nu)$
in the anarchic triplet model, the anarchic bidoublet model and the bidoublet model with $U(2)^3$ left-handed compositeness.
In all plots, the gray regions correspond to the experimental $1\sigma$ ranges. In the case of $\text{BR}(K_L\to\mu^+\mu^-)_\text{SD}$, the horizontal gray line shows the bound in eq.~(\ref{eq:Kmmbound}). In the upper left plot, the diagonal line shows the model-independent Grossman-Nir bound \cite{Grossman:1997sk}.
The black cross in each plot shows the SM central values and uncertainties. All the new physics points are obtained for central values of the theory parameters, so are understood to carry at least the relative uncertainty of the SM point.

In the case of the anarchic triplet model, represented by the yellow points in figure~\ref{fig:BR_all}, the following observations can be made.
\begin{itemize}
\item $K^+\to\pi^+\nu\bar\nu$ and $K_L\to\mu^+\mu^-$ already cut into the model's parameter space and $\text{BR}(K_L\to\pi^0\nu\bar\nu)$ can be enhanced to five times its SM value.
\item $\text{BR}(K^+\to\pi^+\nu\bar\nu)$ and $\text{BR}(K_L\to\mu^+\mu^-)_\text{SD}$ are anticorrelated. This is because only the right-handed flavour-changing $Z$ coupling contributes. An analogous effect has been observed in a 5D model with a similar setup \cite{Blanke:2008yr}.
\item The bound on $K_L\to\mu^+\mu^-$ excludes a suppression of $\text{BR}(K^+\to\pi^+\nu\bar\nu)$ by more than 50\%.
\item Both $B_s\to\mu^+\mu^-$ and $B_d\to\mu^+\mu^-$ can be suppressed or enhanced by up to 50\%.
The precision of the $B_s\to\mu^+\mu^-$ measurement is now entering the region accessible in the model.
\item $\text{BR}(B_s\to\mu^+\mu^-)$ and $\text{BR}(B\to K^*\mu^+\mu^-)$ are perfectly correlated. From equations (\ref{eq:Bsmumu}) and (\ref{eq:BKsmumu}) one can see that such correlation is present if only left-handed or only right-handed couplings (as in this case) contribute, while deviations are possible if both left- and right-handed couplings are present.
\item The CP asymmetry $A_9$ in $B\to K^*\mu^+\mu^-$ at high $q^2$ can reach at most 5\%; the CP-averaged observable $S_3$ can deviate by $\pm10\%$ from its SM expectation. At present, the theory uncertainty due to form factors is too large to constrain the model.
\item $\text{BR}(B\to K \nu\bar\nu)$ and $\text{BR}(B\to K^* \nu\bar\nu)$ are anticorrelated, again because only the right-handed flavour-changing $Z$ coupling contributes (cf.\ \cite{Altmannshofer:2009ma,Buras:2010pz,Straub:2010ih,Buras:2012jb}).
\end{itemize}
Note that although the experimental measurement of the $B\to K^*\mu^+\mu^-$ branching ratio is already quite precise, the large SM uncertainty limits its constraining power at present. The situation will change once 
high-$q^2$ form factors will be computed on the lattice.

The red points in figure~\ref{fig:BR_all} show the analogous correlations in the anarchic bidoublet model. One can make the following observations.
\begin{itemize}
\item The effects in $K^+\to\pi^+\nu\bar\nu$ and $K_L\to\pi^0\nu\bar\nu$ are almost unconstrained by experiment as of yet, but enhancements by factors of 2 are possible.
\item $\text{BR}(K^+\to\pi^+\nu\bar\nu)$ and $\text{BR}(K_L\to\mu^+\mu^-)_\text{SD}$ are now perfectly correlated rather than anticorrelated since only the left-handed flavour-changing $Z$ coupling contributes.
\item The measurement of $B_s\to\mu^+\mu^-$ already constrains the model signifcantly, while effects in $B_d\to\mu^+\mu^-$ are at most $\pm50\%$, not yet constrained by experiment. The two decays are uncorrelated.
\item $\text{BR}(B_s\to\mu^+\mu^-)$ vs.\ $\text{BR}(B\to K^*\mu^+\mu^-)$ shows almost the same correlation as in the triplet model. This is because now, only left-handed couplings contribute.
\item The observables $S_3$ and $A_9$ in $B\to K^*\mu^+\mu^-$ are SM-like since there are no right-handed flavour-changing $Z$ couplings.
\item Also $\text{BR}(B\to K \nu\bar\nu)$ and $\text{BR}(B\to K^* \nu\bar\nu)$ are now correlated rather than anticorrelated, because the effects are due to left-handed couplings.
\end{itemize}

Finally, the dark blue points in figure~\ref{fig:BR_all} show the correlations for the $U(2)^3$ bidoublet model with left-handed compositeness. There are two main differences with respect to the anarchic bidoublet model.
\begin{itemize}
\item $K^+\to\pi^+\nu\bar\nu$ and $K_L\to\pi^0\nu\bar\nu$ are perfectly correlated since the amplitude is aligned in phase with the SM.
\item Also $B_s\to\mu^+\mu^-$ and $B_d\to\mu^+\mu^-$ are perfectly correlated since $b\to d$ and $b\to s$ amplitudes are universal, up to the same CKM factors as in the SM.
\end{itemize}
Both these correlations are universal predictions of models with Minimal Flavour Violation (MFV) \cite{Buras:2000dm,D'Ambrosio:2002ex} and of models with a minimally broken $U(2)^3$ symmetry (as in this case).

\section{Conclusions and Outlook}\label{sec:conc}

In models with partial compositeness, flavour-changing $Z$ couplings can lead to visible new physics effects in rare $K$ and $B$ decays. The chirality, phase and size of these flavour-changing couplings depend on the one hand on the mechanism imposed to keep the flavour-conserving $Z\to b\bar b$ coupling under control and on the other hand on the presence or absence of a flavour symmetry imposed to ameliorate the $\epsilon_K$ problem. In this paper, these effects have been analyzed numerically in three scenarios, two flavour-anarchic ones with different fermion content (dubbed triplet and bidoublet models) and a model with a $U(2)^3$ flavour symmetry, minimally broken only by the composite-elementary mixings of right-handed quarks. In all cases, a two-site Lagrangian was used with some simplifying assumptions, e.g.\ a common mass and coupling for all the vector resonances. A new method was used to sample the highly-dimensional parameter space of these models.

The main result of this study is figure~\ref{fig:BR_all}. It shows the size of and correlations between the most relevant rare $K$ and $B$ decays. The observables that have been chosen are not sensitive to magnetic dipole operators that are also generated in these models, so they can be considered clean probes of flavour-changing $Z$ couplings.

Interestingly, in all the scenarios, all the branching ratios considered can exhibit visible effects at current and planned experiments such as LHCb, Belle-II, KOTO and NA62. Particularly large effects, already constrained by existing experimental results, are possible in $K$ decays in the anarchic triplet model and in $B$ decays in the case of the bidoublet models. Clear-cut correlations between various observables can be used to distinguish the three models on the basis of the pattern of observed deviations from the SM expectations.

This study could be extended in various ways. Going beyond the simple two-site picture and adopting more complete frameworks, e.g.\ in a 5D picture, additional relations between the parameters could modify the maximum size of the effects found here.  In addition, a number of potentially relevant constraints has not been taken into account in this analysis, namely ones related to loop-induced observables. The rationale is that loop effects are more dependent on the details of the theory, e.g.\ the localization of the Higgs in the extra dimension in a 5D picture (see e.g.\ \cite{Casagrande:2010si,Carena:2012fk,Delaunay:2012cz}). In concrete frameworks however, such effects can be calculable. The most important loop-induced quantities are the $T$ parameter and magnetic dipole operators contributing to FCNCs and electric dipole moments. Their study is left to a future publication.

\section*{Acknowledgments}

I thank
Paul Archer,
Andrzej Buras,
Dario Buttazzo,
Sandro Ca\-sa\-gran\-de,
Raoul Malm,
Filippo Sala,
and
Andrea Tesi
for useful discussions and
Riccardo Barbieri
and
Matthias Neubert
for valuable comments on the manuscript.
This research is supported by the Advanced Grant EFT4LHC of the European Research Council (ERC), and the Cluster of Excellence {\em Precision Physics, Fundamental Interactions and Structure of Matter\/} (PRISMA -- EXC 1098).

\appendix
\section{Details on the numerical analysis}\label{sec:mcmc}

\subsection{Scanning procedure}

The numerical analysis of partial compositeness models with anarchic Yukawa couplings is challenging since the quark masses and mixings depend on a large number of free parameters, making a brute-force scan of the highly-dimensional parameter space unfeasible. The conventional approach to tackle this problem is to exploit approximate analytical relations between the Yukawas, degrees of compositeness and the masses and mixings \cite{Casagrande:2008hr,Blanke:2008zb,Bauer:2009cf}. This approach has several drawbacks. First, it is difficult to obtain the correct CKM phase, so that a partially brute-force approach is necessary. Second, the approximate analytical expressions become very complicated if the left-handed quark doublet couples to two different fields, as is the case in the bidoublet model. Third, once a set of viable parameter points is obtained, imposing additional constraints, like $Z\to b\bar b$ or $\Delta F=2$ observables (especially $|\epsilon_K|$), one looses a large fractions of the points.

To solve these problems, this study uses an alternative approach based on Markov Chain Monte Carlo (MCMC) with the Metropolis-Hastings algorithm. This method is a very simple tool to sample a highly-dimensional probability distribution, given a likelihood function and a prior distribution for the input parameters. While it is frequently used for Bayesian parameter estimation, here a different property is exploited: the algorithm automatically samples the region in parameter space where the likelihood is high, but (in the stationary limit, i.e.\ with infinitely long chains) is supposed to sample the entire relevant parameter space.

The procedure is as follows. One defines the likelihood as $L=e^{-\chi^2(\theta)/2}$, where the $\chi^2$ function depends on all the model input parameters $\theta$ (including the real and imaginary parts of the Yukawa matrices, the composite-elementary mixings, etc.) and contains the experimentally measured quark masses and CKM angles. One now starts a chain with step size chosen to obtain an acceptance rate around $20\%$. That is, after a burn-in period, out of 5 generated points, 1 point is viable. The chain has to be long enough to obtain an acceptable coverage of the parameter space. Several chains with random initial values can be combined to make sure disjoint minima are not missed.

One can now go one step further and include additional constraints in the $\chi^2$. This was done in the numerical analysis of section~\ref{sec:numerics} for the $Z\to b\bar b$ partial width and the $Z$ hadronic width as well as for the $\Delta F=2$ observables $\Delta M_d$, $\Delta M_s/\Delta M_d$, $S_{\psi K_S}$, $\phi_s$ and $\epsilon_K$. Apart from drastically reducing the number of points to be generated -- since constraints are automatically fulfilled -- this has an additional advantage: in the conventional approach, the CKM angles and phase to be used in the first step are always the same, while the $\Delta F=2$ processes calculated in the second step actually change the preferred CKM parameters. If $\Delta F=2$ constraints are included alongside CKM angles obtained from {\em tree-level} processes, the impact of new physics effects in loops on the CKM determination is automatically taken into account. For example, a sizable constructive new physics contribution to the $B_d$ mixing phase leads to a 
lower value of $\bar\rho$ in 
the determined point, so that $S_{\psi K_S}$ agrees with the experimental value. 

While the acceptance rate can be kept high by adjusting the step size, the downside of including additional constraints is an increase in the required burn-in time as well as a reduced coverage of the parameter space for a given chain length. Consequently, one needs a larger number of points to obtain a representative sample. For each of the plots in section~\ref{sec:numerics}, approximately $10^6$ viable points, consisting of 5-10 chains, were used.

It should be stressed again that the only purpose of the MCMC approach is to sample the regions in parameter space which have the right quark masses and mixings and which fulfill the flavour cosntraints. The purpose is {\em not} to obtain posterior probability distributions for the input parameters. This would be a futile task since the number of input parameters far exceeds the number of constraints. The dependence on the choice of priors and the limited coverage are thus much less problematic than they would be in the case of a probabilistic analysis.

\subsection{Input parameters and constraints}

\begin{table*}
 \begin{center}
\begin{tabular}{llllll}
\hline
$\lambda$ & $0.22546(22)$ &\cite{Bevan:2010gi}& $f_K$  & $(156.1\pm1.1)$ MeV & \cite{Laiho:2009eu}\\
$A$ & $0.807\pm0.020$ &\cite{Bevan:2010gi}& $\hat B_K$ & $0.764\pm0.010$ &\cite{Laiho:2009eu} \\
$\bar\rho$ & $0.077\pm0.067$ &\cite{Bevan:2010gi}& $\kappa_\epsilon$ & $0.94\pm0.02$ & \cite{Buras:2010pza}\\
$\bar\eta$ & $0.384\pm0.061$ &\cite{Bevan:2010gi}& $f_{B_s}\sqrt{\hat B_s}$  & $(279\pm15)$ MeV &\cite{Laiho:2009eu}\\
$\gamma$ & $(75.5\pm10.5)^\circ$ &\cite{Bevan:2010gi}& $\xi$ & $1.237\pm0.032$ &\cite{Laiho:2009eu}\\
$|\epsilon_K|$ & $(2.229\pm0.010)\times10^{-3}$ &\cite{Nakamura:2010zzi} &$\eta_{tt}$&$0.5765(65)$&\cite{Buras:1990fn}\\
$S_{\psi K_S}$ & $0.679\pm0.020$ &\cite{Asner:2010qj} &$\eta_{ct}$&$0.496(47)$&\cite{Brod:2010mj}\\
$\Delta M_d$ & $(0.507\pm0.004)\,\text{ps}^{-1}$ &\cite{Asner:2010qj} &$\eta_{cc}$&$1.87(76)$&\cite{Brod:2011ty}\\
$\Delta M_s/\Delta M_d$ & $(35.05\pm0.42)\,\text{ps}^{-1}$ &\cite{Abulencia:2006ze,Asner:2010qj} &&&\\
$\phi_s$ & $-0.002\pm0.087$ &\cite{LHCb-CONF-2012-002} &&&\\
\hline
 \end{tabular}
 \end{center}
\caption{Experimental values and hadronic parameters used for the $\Delta F=2$ constraints.}
\label{tab:input}
\end{table*}

For the inclusion of the $\Delta F=2$ constraints, the Wilson coefficients of the operators
\begin{align}
Q_V^{LL} &= (\bar d^i_L\gamma^\mu d^j_L)(\bar d^i_L\gamma^\mu d^j_L) \,,
&
Q_V^{RR} &= (\bar d^i_R\gamma^\mu d^j_R)(\bar d^i_R\gamma^\mu d^j_R) \,,
\\
Q_V^{LR} &= (\bar d^i_L\gamma^\mu d^j_L)(\bar d^i_R\gamma^\mu d^j_R) \,,
&
Q_S^{LR} &= (\bar d^i_R d^j_L)(\bar d^i_L d^j_R) \,,
\end{align}
with $ij=21,31,32$ are evaluated at tree level, including contributions from neutral coloured and uncoloured vector resonances. The renormalization group evolution from the scale $m_\rho$ to the appropriate low-energy scale follows ref.~\cite{Buras:2001ra}. The input parameters used are collected in table~\ref{tab:input}. The CKM Wolfenstein parameters correspond to a fit to tree-level observables only, since the loop-induced observables are modified by new physics and are taken into account by the scanning procedure as discussed in the previous section. The uncertainties of the hadronic input parameters in the right-hand column are used to determine the overall theory uncertainty on each observable, which then enters the $\chi^2$ function, added in quadrature with the experimental uncertainty.
All uncertainties are assumed to be Gaussian.

For the two most significant constraints, slightly looser uncertainties where assumed.
For $|\epsilon_K|$, the theory uncertainty was assumed to be at least 30\% of the central value of the SM prediction.
In the case of $R_b$, the partial decay width of the $Z$ to $b$ quarks, fits currently show a $2.5\sigma$ deviation from the SM prediction \cite{Baak:2012kk}, which is always worsened (or stays the same) in the models at hand \cite{Barbieri:2012tu}. In order not to be too restrictive, $R_b$ was treated in the $\chi^2$ as if the experimental central value coincided with the SM central value, effectively requiring $R_b$ to be within $1\sigma$ of the SM value. Otherwise even SM-like points would be disfavoured. The same was done for the hadronic width $R_h$, relevant in the $U(2)^3$ case.

\bibliographystyle{JHEP}
\bibliography{zpenguins}

\end{document}